\documentclass[twocolumn,showpacs,preprintnumbers,amsmath,amssymb]{revtex4}
\usepackage{natbib}
\usepackage{euscript}
\usepackage{amsmath}
\usepackage{amssymb}
\usepackage{graphicx}

\begin{document}

\title{Instabilities in the two-dimensional cubic nonlinear Schr\"odinger equation}
\author{John D. Carter}
\affiliation{Seattle University\\Mathematics Department\\Seattle, WA 98122}
\email{carterj1@seattleu.edu}
\author{Harvey Segur}
\affiliation{University of Colorado at Boulder\\Department of Applied
  Mathematics\\Boulder, CO 80309}
\email{segur@colorado.edu}

\date{\today}

\begin{abstract}

The two-dimensional cubic nonlinear Schr\"odinger equation (NLS) can be 
used as a model of phenomena in physical systems ranging from waves on 
deep water to pulses in optical fibers.  In this paper, we establish
that every one-dimensional traveling wave solution of NLS with trivial
phase is unstable with respect to some infinitesimal
perturbation with two-dimensional structure.  If the
coefficients of the linear dispersion terms have the same sign then
the only unstable perturbations have transverse wavelength longer than
a well-defined cut-off.  If the coefficients of the
linear dispersion terms have opposite signs, then there is no such
cut-off and as the wavelength decreases, the maximum growth rate
approaches a well-defined limit.

\end{abstract}

\pacs{42.65.Sf, 92.10.Hm, 47.35.+i, 02.30.Jr}

\maketitle

\section{Introduction}

The two-dimensional cubic nonlinear Schr\"odinger equation (NLS) is
given by
\begin{equation}
i\psi_t+\alpha \psi_{xx}+\beta \psi_{yy}+\gamma|\psi|^2\psi=0,
\label{NLSE}
\end{equation}
where $\psi=\psi(x,y,t)$ is a complex-valued function,
and $\alpha$, $\beta$
and $\gamma$ are real constants. Among many other situations, NLS arises as an
approximate model of the evolution of a nearly monochromatic wave of
small amplitude in pulse propagation along optical
fibers~\cite{opticswaveguides} where
$\alpha\beta>0$, in gravity waves on deep water~\cite{BR}~\cite{DS}
where $\alpha\beta<0$ and in Langmuir waves in a plasma~\cite{Pecseli}
where $\alpha\beta>0$.  As a description of a
superfluid~\cite{Donnelly}, NLS is
known as the Gross-Pitaevskii equation
\cite{Gross}~\cite{Pitaevskii} with
$\alpha\beta>0$.  Sulem and Sulem~\cite{sulemsulem} examine NLS in detail. 

NLS admits a large class of one-dimensional traveling wave
solutions of the form 
\begin{equation}
\psi(x,y,t)=\phi(ax+by-st)\mbox{e}^{i\lambda t+ia\bar{s}x+ib\bar{s}y+i\eta},
\label{completeform}
\end{equation}
where $\phi$ is a real-valued function,
$\bar{s}=s/(2\alpha a^2+2\beta b^2)$ and $a$, $b$, $s$,
$\lambda$ and $\eta$ are real parameters.  By making use of the symmetries of NLS~\cite{sulemsulem}, all solutions of this form can be considered by studying the simplified form
\begin{equation}
\psi(x,y,t)=\phi(z)\mbox{e}^{i\kappa x+i\lambda t},
\label{simpleform}
\end{equation}
where $\phi$ is a real-valued function, $z=x-2\alpha\kappa t$, and $\kappa$
and $\lambda$ are real parameters.

If $\alpha\gamma>0$, then NLS admits the following two solutions
of the form (\ref{simpleform})
\begin{equation}
\phi(z)=\sqrt{2\frac{\alpha}{\gamma}}\;k\;\mbox{cn}(z,k),\;\;\text{with}\;\;\lambda=\alpha (2k^2-1-\kappa^2),
\label{cnNLSEsolution}
\end{equation}
\begin{equation}
\phi(z)=\sqrt{2\frac{\alpha}{\gamma}}\;\mbox{dn}(z,k),\;\;\text{with}\;\;\lambda=\alpha (2-k^2-\kappa^2).
\label{dnNLSEsolution}
\end{equation}
If $\alpha\gamma<0$, then NLS admits the following solution
\begin{equation}
\phi(z)=\sqrt{-2\frac{\alpha}{\gamma}}\;k\;\mbox{sn}(z,k),\;\text{with}\;\lambda=-\alpha(1+k^2+\kappa^2).
\label{snNLSEsolution}
\end{equation}
Here $k\in[0,1]$ is a free parameter known as the elliptic modulus and
$\mbox{cn}(\cdot,k)$, $\mbox{dn}(\cdot,k)$, and $\mbox{sn}(\cdot,k)$
are Jacobi elliptic functions.  Byrd and Friedman~\cite{byrd} provide
a complete review of elliptic functions.  If $k<1$, then each function
$\phi(z)$ is periodic.  As $k\rightarrow 1$, the period of each increases without bound, and $\phi(z)$ limits to an appropriate hyperbolic function, which we call a ``solitary wave.''

These solutions, plus the ``Stokes' wave'' (plane wave)
\begin{equation}
\psi(x,t)=A\mbox{e}^{i\kappa x-i(\alpha\kappa^2-\gamma|A|^2)t},
\end{equation}
comprise the entire class of bounded traveling wave solutions of NLS with trivial phase~\cite{carr1}.  Davey and Stewartson~\cite{DS} show that a Stokes' wave is unstable unless either $\alpha\beta\gamma=0$, or $\alpha\beta>0$ and $\alpha\gamma<0$.  In the remainder of this paper, we concentrate on (\ref{cnNLSEsolution}), (\ref{dnNLSEsolution}) and (\ref{snNLSEsolution}), and on their instabilities.

Zakharov and Rubenchik~\cite{zr} establish that
(\ref{cnNLSEsolution}) and (\ref{dnNLSEsolution}) with $k=1$ are
unstable with respect to long-wave transverse perturbations.
Pelinovsky~\cite{pel} reviews the stability of solitary wave solutions
of NLS with $\alpha\beta<0$ and $\alpha\gamma>0$, and presents an
analytical expression for the growth rate of the instability near a
cut-off.  Extensive reviews of the stability of solitary wave
solutions are given in~\cite{krz,rr,kivpel}.  The periodic
problem has not been studied in as much detail, though Martin,
Yuen and Saffman~\cite{MYS} examine numerically the stability of the solution given in (\ref{dnNLSEsolution}) for a range of parameters.  

We present four main results in this paper.  First, every
one-dimensional traveling wave with trivial phase is
unstable with respect to some infinitesimal perturbation with
two-dimensional structure.  For \emph{all} choices of the parameters,
there are unstable perturbations with long transverse wavelength.
This generalizes the result of ~\cite{zr}. 

Second, if $\alpha\beta>0$, then the only unstable perturbations have transverse wavelength longer than a well-defined cut-off.

Third, if $\alpha\beta<0$, then there is no such cut-off.  There are unstable perturbations with arbitrarily short wavelengths in both transverse and longitudinal directions.  These short wavelength instabilities seem to have been overlooked in previous analyses.

Fourth, for $\alpha\beta<0$, the unstable perturbations with short wavelength have transverse wavenumbers that are confined to narrower and narrower intervals as the transverse wavenumber grows without bound.  In these unstable intervals, as the transverse wavenumber grows without bound, the maximum growth rate approaches a well-defined limit.  As $k\rightarrow 1$, this limiting growth rate tends to zero if $\alpha\gamma>0$, and to a finite non-zero limit if $\alpha\gamma<0$.

\section{Stability Analysis}

We consider perturbed solutions, $\psi_{_p}=\psi_{_p}(x,y,t)$, with the following structure
\begin{equation}
\psi_{_p}=(\phi(z)+\epsilon u(x,y,t)+i \epsilon
v(x,y,t)+O(\epsilon^2))\mbox{e}^{i\kappa x+i\lambda t},
\label{pert}
\end{equation}
where $u(x,y,t)$ and $v(x,y,t)$ are real-valued functions, $\epsilon$ is a
small real parameter, $z=x-2\alpha\kappa t$, and $\phi(z)\mbox{e}^{i\kappa x+i\lambda t}$ is one of the
solutions presented in the previous section.  Substituting (\ref{pert}) into (\ref{NLSE}),
linearizing and separating into real and imaginary parts gives
\begin{subequations}
\begin{equation}
-(\alpha\kappa^2+\lambda) u+3\gamma\phi^2u+\beta
u_{yy}+\alpha u_{xx}=v_t,\label{reale}
\end{equation}
\begin{equation}
-(\alpha\kappa^2+\lambda) v+\gamma\phi^2v+\beta v_{yy}+\alpha v_{xx}=-u_t.
\label{image}
\end{equation}
\label{evaluepde}
\end{subequations}
Without loss of generality, assume that $u(x,y,t)$ and $v(x,y,t)$ have
the forms
\begin{subequations}
\begin{equation}
u(x,y,t)=U(z,\rho)\mbox{e}^{i\rho y-\Omega t}+c.c.,
\end{equation}
\begin{equation}
v(x,y,t)=V(z,\rho)\mbox{e}^{i\rho y-\Omega t}+c.c.,
\end{equation}
\label{uvtrivial}
\end{subequations}
where $\rho$ is a real constant, $\Omega$ is a complex constant, 
$U$ and $V$ are complex-valued functions and $c.c.$ denotes
complex conjugate.  This leads to
\begin{subequations}
\begin{equation}
(\alpha\kappa^2+\lambda) U-3\gamma\phi^2U
+\beta\rho^2U-\alpha \partial_z^2 U=\Omega V,\label{evprob1}
\end{equation}
\begin{equation}
(\alpha\kappa^2+\lambda) V-\gamma\phi^2V+\beta\rho^2 V-\alpha\partial_z^2
V=-\Omega U.
\label{evprob2}
\end{equation}
\label{evprob}
\end{subequations}
These are the central equations in this paper.
We assume that $U$ and $V$ are periodic with the same period as
$\phi$.  More general boundary conditions are discussed
in~\cite{mythesis}.   
Instability occurs if (\ref{evprob}) admits a periodic solution with
$Re(\Omega)<0$.  Without loss of generality, for the remainder of this
paper we assume $\kappa=0$ by redefining $\lambda$. 

In Sections \ref{smallrhosection} and \ref{largerhosection}, we
examine (\ref{evprob}) using small-$\rho$ and large-$\rho$
asymptotic analyses respectively.  In Section \ref{monodromysection},
we present results from a numerical study in which (\ref{evprob})
was solved for a wide range of $\rho$ values.  

\section{Small-$\rho$ limit}
\label{smallrhosection}

Generalizing the work in~\cite{zr}, we assume that for fixed $k$ and for fixed small $\rho$, (\ref{evprob}) admits solutions of the form
\begin{subequations}
\begin{equation}
U\sim u_0(z)+\rho u_1(z)+\rho^2 u_2(z)+\cdots,
\label{U}
\end{equation}
\begin{equation}
V\sim v_0(z)+\rho v_1(z)+\rho^2 v_2(z)+\cdots,
\label{V}
\end{equation}
\begin{equation}
\Omega^2\sim\rho^2\omega_1+\rho^3\omega_2+\cdots,
\label{O}
\end{equation}
\end{subequations}
where the $\omega_j$ are complex constants and the $u_j$ and $v_j$ are
complex-valued periodic functions with the same period as $\phi$.

\begin{figure}
\centerline{\includegraphics[width=6cm]{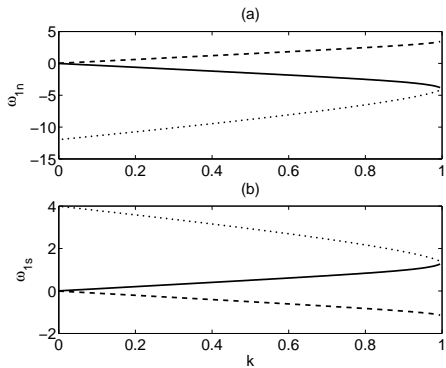}}
\caption{Plots of $\omega_{1n}$ and $\omega_{1s}$ versus $k$ in (a)
  and (b) respectively.  The solid line corresponds to
  (\ref{cnNLSEsolution}), the dotted line corresponds to
(\ref{dnNLSEsolution}) and the dashed line corresponds to
(\ref{snNLSEsolution}).}
\label{smallrhoplots}
\end{figure}

This assumption leads to the ``neck'' mode
\begin{subequations}
\begin{equation}
U_n(z,\rho)=O(\rho),
\label{Un}
\end{equation}
\begin{equation}
V_n(z,\rho)= \phi+O(\rho),
\label{Vn}
\end{equation}
\begin{equation}
\Omega_n^2= -\alpha\beta \rho^2\omega_{1n}+O(\rho^3),
\label{omegan}
\end{equation}
\end{subequations}
and the ``snake'' mode
\begin{subequations}
\begin{equation}
U_s(z,\rho)= \frac{d\phi}{dz}+O(\rho),
\label{Us}
\end{equation}
\begin{equation}
V_s(z,\rho)=O(\rho),
\label{Vs}
\end{equation}
\begin{equation}
\Omega_s^2= -\alpha\beta \rho^2\omega_{1s}
+O(\rho^3),
\label{omegas}
\end{equation}
\end{subequations}
where $\omega_{1n}$ and
$\omega_{1s}$ are functions of the elliptic modulus of the unperturbed
solution.  Complicated but exact expressions for $\omega_{1n}$ and
$\omega_{1s}$ are derived in~\cite{mythesis}.  The final results are
presented in Fig.\ \ref{smallrhoplots}, where we plot $\omega_{1n}$
and $\omega_{1s}$, the growth rates, versus $k$ for (\ref{cnNLSEsolution}), (\ref{dnNLSEsolution}) and
(\ref{snNLSEsolution}).

These plots establish that $\omega_{1n}<0$ for (\ref{cnNLSEsolution})
and (\ref{dnNLSEsolution}) and $\omega_{1s}<0$ for
(\ref{snNLSEsolution}).  Therefore, if $\alpha\beta>0$,
(\ref{cnNLSEsolution}) and (\ref{dnNLSEsolution}) are unstable with respect to
long-wave transverse perturbations corresponding to the neck mode
and (\ref{snNLSEsolution}) is unstable with respect to long-wave
transverse perturbations corresponding to the the snake mode.

These plots also establish that $\omega_{1s}>0$ for
(\ref{cnNLSEsolution}) and (\ref{dnNLSEsolution}) and $\omega_{1n}>0$
for (\ref{snNLSEsolution}).  Therefore, if $\alpha\beta<0$, 
(\ref{cnNLSEsolution}) and (\ref{dnNLSEsolution}) are unstable with
respect to the snake mode and (\ref{snNLSEsolution}) is unstable with
respect to the neck mode.   

It follows from these results that a trivial-phase solution of NLS is
unstable to a growing neck mode if $\beta\gamma>0$, and to a growing
snake mode if $\beta\gamma<0$. 

\section{Large-$\rho$ limit}
\label{largerhosection}

\begin{figure}[floatfix]
\centerline{\includegraphics[width=6cm]{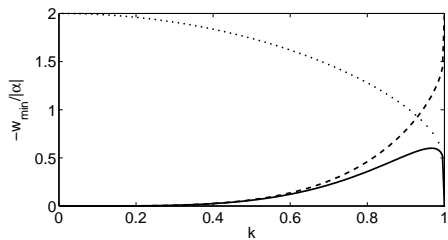}}
\caption{Plots of $-w_{min}/|\alpha|$ versus $k$.  The solid line
  corresponds to (\ref{cnNLSEsolution}), the dotted line corresponds
  to (\ref{dnNLSEsolution}) and the dashed line corresponds to
  (\ref{snNLSEsolution}).} 
\label{largerho}
\end{figure}

If $\alpha\beta>0$ and $\rho$ is chosen to be large enough to satisfy
\begin{equation}
\rho^2>5|\frac{\alpha}{\beta}|,
\end{equation}
then the two operators on the left side of (\ref{evprob}) have the
same sign, so $\Omega^2<0$.  Therefore, there is no large-$\rho$
instability if $\alpha\beta>0$.

If $\alpha\beta<0$ and $\rho$ is large, then one can show that there is no instability unless $\Omega=O(1)$.  Therefore, we assume
\begin{subequations}
\begin{equation}
U\sim\zeta_1(\mu z)+\rho^{-2}\zeta_2(\mu z)+\cdots,
\label{Uform}
\end{equation}
\begin{equation}
V\sim\xi_1(\mu z)+\rho^{-2}\xi_2(\mu z)+\cdots,
\label{Vform}\end{equation}
\begin{equation}
\Omega\sim w_1+\rho^{-2}w_2+\cdots,
\label{Omegaform}
\end{equation}
\begin{equation}
\alpha\mu^2=-\beta\rho^2+\nu+O(\rho^{-2}),
\label{mudef}
\end{equation}
\label{largeforms}
\end{subequations}
where $\nu$ is a real constant, the $w_j$ are complex constants, and the
$\zeta_j$ and $\xi_j$ are complex-valued periodic functions with the
same period as $\phi$.

Substituting (\ref{largeforms}) into (\ref{evprob}), one finds at
leading order 
\begin{subequations}
\begin{equation}
U\sim\zeta_{11}\sin(\mu z+z_0) + O(\rho^{-2}),
\end{equation}
\begin{equation}
V\sim\xi_{11}\sin(\mu z+z_0) + O(\rho^{-2}),
\end{equation}
\label{UVlargerho}
\end{subequations}
where $\zeta_{11}$, $\xi_{11}$ and $z_0$ are constants.  Requiring $U$ and $V$ to have the same period as $\phi(z)$ forces $\mu$ to take on discrete values: $\mu=2\pi N/L$, where $L$ is the period of $\phi(z)$, and $N$ is an integer.  To satisfy (\ref{mudef}), $N\gg 1$.  At the next order in $\rho$, solutions are periodic only if 
\begin{equation}
w_1=\pm\sqrt{\gamma f-\lambda+\nu}\sqrt{\lambda-3\gamma f-\nu},
\label{omega1}
\end{equation}
where $f$ is the Fourier coefficient (in $\sin(\mu z+z_0)$) of
$\phi^2(z)\sin(\mu z+z_0)$ and $\nu$ is $O(1)$ but otherwise
arbitrary.  Minimizing the negative root in (\ref{omega1}) with
respect to $\nu$ leads to 
\begin{equation}
w_{min}=-|\gamma f|,
\label{omegaAmax}
\end{equation}
when $\nu=(\lambda-2\gamma f)$.  Then (\ref{mudef}) defines $\rho_{_N}$, the value of $\rho$ at which the Nth unstable mode achieves its maximum growth rate:
\begin{equation}
-\beta\rho^2_{_N}=\alpha(2\pi N/L)^2+2\gamma f-\lambda.
\end{equation}
We also find how for $\rho$ can deviate from $\rho_{_N}$ before
$\Omega^2$ becomes negative
\begin{equation}
\delta\rho_{_N}\sim|\gamma|f/(2|\beta|\rho_{_N})=O(1/N),\quad\text{for}\quad
N\gg1.
\end{equation}

Analytic expressions for the $f$ corresponding to the solutions given
in (\ref{cnNLSEsolution}), (\ref{dnNLSEsolution}) and
(\ref{snNLSEsolution}) are not known.  But, in the
large-$\rho$ limit, the Riemann-Lebesgue
Lemma~\cite{GuentherLee} can be used to determine approximate
expressions.  As $\rho\rightarrow\infty$, for the solution given in
(\ref{cnNLSEsolution}), 
\begin{equation} 
f\sim\frac{1}{2K(k)}\big{|}\frac{\alpha}{\gamma}\big{|}\Big{(}E(\mbox{am}(4K(k)))+4(k^2-1)K(k)\Big{)},
\end{equation}
for the solution given in (\ref{dnNLSEsolution}),
\begin{equation}
f\sim \frac{1}{K(k)}\big{|}\frac{\alpha}{\gamma}\big{|}\Big{(}E(\mbox{am}(2K(k)))\Big{)},
\end{equation}
and for the solution given in (\ref{snNLSEsolution}),
\begin{equation}
f\sim \frac{1}{2K(k)}\big{|}\frac{\alpha}{\gamma}\big{|}\Big{(}4K(k)-E(\mbox{am}(4K(k)))\Big{)}.
\end{equation}
In each of these expressions, $\mbox{am}(\cdot)$ gives the Jacobi amplitude and $K(\cdot)$ and $E(\cdot)$ are the complete elliptic integrals of the first and second kind respectively~\cite{byrd}.

Plots of $-w_{min}/|\alpha|$, a growth rate, versus $k$ are given in Fig.\
\ref{largerho}.  This argument establishes that all finite-period
one-dimensional trivial phase solutions are unstable with respect to
arbitrarily short-wavelength transverse perturbations if
$\alpha\beta<0$, and that the growth rate of the instability remains
bounded as $\rho\rightarrow\infty$.  

Note that as $k\rightarrow 1$, $\phi(z)$ in (\ref{snNLSEsolution}) 
approaches a hyperbolic tangent and the growth rate approaches that of the Stokes' wave with an amplitude of
$\sqrt{-2\alpha/\gamma}$.  This establishes that there
are an infinite number of of unstable branches if $\alpha\beta<0$ and
$\alpha\gamma<0$. 

Also note that as $k\rightarrow 1$, $\phi(z)$ in both (\ref{cnNLSEsolution}) and (\ref{dnNLSEsolution}) approaches a hyperbolic secant, and the corresponding growth rate limits to zero.  This establishes that there is no large-$\rho$ instability in the solitary wave limit if $\alpha\gamma>0$. 

\section{Monodromy}
\label{monodromysection}

\begin{figure}[floatfix]
\centerline{\includegraphics[width=7cm]{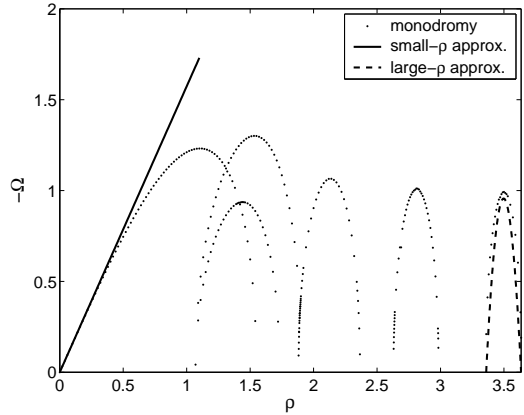}}
\caption{Plots of $-\Omega$ versus $\rho$ corresponding to
(\ref{cnNLSEsolution}) with\\ $k=\sqrt{0.8}$ and
$-\alpha=\beta=\gamma=1$.  See text for a description.} 
\label{snmppdata}
\end{figure}

\begin{figure}[floatfix]
\centerline{\includegraphics[width=7cm]{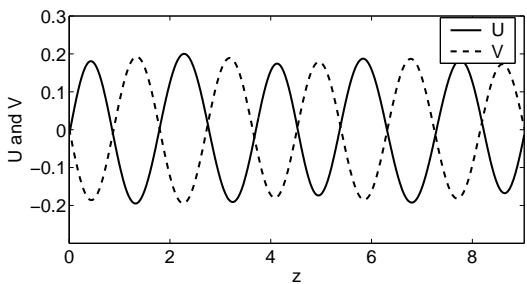}}
\caption{Plots of $U$ and $V$ versus $z$ corresponding to $(\rho=3.5,
\Omega=-0.99,N=5)$.} 
\label{instabsplots}
\end{figure}

The system of equations in (\ref{evprob}) is Hamiltonian in $z$, with periodic boundary conditions.  The coordinates on the phase space are $p_1=dU/dz$, $p_2=-dV/dz$, $q_1=U$ and $q_2=V$.  The Hamiltonian is
\begin{eqnarray}
H=\frac{1}{2}(p_1^2-p_2^2)-\frac{1}{2\alpha}(\lambda+\beta\rho^2-3\gamma\phi^2(x))q_1^2\nonumber
\\+\frac{1}{2\alpha}(\lambda+\beta\rho^2-\gamma\phi^2)q_2^2+\frac{\Omega}{\alpha}q_1q_2.
\label{evham}
\end{eqnarray}

Such a Hamiltonian system necessarily has a monodromy structure with
invariants~\cite{meyerhall}.  We used this structure to identify the
periodic solutions of (\ref{evprob}) by numerically integrating
(\ref{evprob}) over one period of $\phi$.   

The growth rates obtained from numerical simulations 
corresponding to (\ref{snNLSEsolution}) with $k=\sqrt{0.8}$ and
$-\alpha=\beta=\gamma=1$ are included in Fig.\
\ref{snmppdata} as dots.  The line is obtained from the small-$\rho$
results.  The dashed curve is obtained from the large-$\rho$ results
with $N=5$.   Each dotted curve corresponds to a different unstable
mode.  A plot of the spatial structure of the mode corresponding to
$(\rho=3.5, \Omega=-0.99, N=5)$ is given in Fig.\ \ref{instabsplots}.   

Figure \ref{snmppdata} demonstrates strong agreement between the
numerical results and the small-$\rho$ analysis when $\rho$ is near zero.
It also demonstrates agreement between the
numerical results and the large-$\rho$ analysis.  

Figure
\ref{instabsplots} demonstrates that the $U$ and $V$ obtained
numerically are similar in form to the $U$ and $V$ obtained in the
large-$\rho$ analysis.  

This work was supported by National Science Foundation Grants
DMS-9731097, DMS-98-10751, and DMS-0139771.  We acknowledge
useful discussions with Bernard Deconinck and Nathan Kutz. 
%\bibliographystyle{apsrev}
%\bibliography{cartersegur}

\end{document}